# Consumers don't play dice, influence of social networks and advertisements

Robert D. Groot

Unilever Research Vlaardingen

PO Box 114, 3130 AC Vlaardingen, The Netherlands

Empirical data of supermarket sales show stylised facts that are similar to stock markets, with a broad (truncated) Lévy distribution of weekly sales differences in the baseline sales [R.D. Groot, *Physica A* 353 (2005) 501]. To investigate the cause of this, the influence of social interactions and advertisements are studied in an agent-based model of consumers in a social network. The influence of network topology was varied by using a small-world network, a random network and a Barabási-Albert network. The degree to which consumers value the opinion of their peers was also varied.

On a small-world and random network we find a phase-transition between an open market and a locked-in market that is similar to condensation in liquids. At the critical point, fluctuations become large and buying behaviour is strongly correlated. However, on the small world network the noise distribution at the critical point is Gaussian, and critical slowing down occurs which is not observed in supermarket sales. On a scale-free network, the model shows a transition between a gas-like phase and a glassy state, but at the transition point the noise amplitude is much larger than what is seen in supermarket sales.

To explore the role of advertisements, a model is studied where imprints are placed on the minds of consumers that ripen when a decision for a product is made. The correct distribution of weekly sales returns follows naturally from this model, as well as the noise amplitude, the correlation time and cross-correlation of sales fluctuations. For particular parameter values, simulated sales correlation shows power law decay in time. The model predicts that social interaction helps to prevent aversion, and that products are viewed more positively when their consumption rate is higher.

# 1  Introduction

Financial markets show fluctuations that closely resemble critical fluctuations as seen in many physical systems. This correspondence has led to many new insights from physics into economy, for instance the Minority Game suggests that markets tend to operate close to a critical point via a mechanism that leads to self-organised criticality [1]. Moreover, it has been established that stock markets fluctuate between an equilibrium state, and a state with a net excess of either buyers of sellers [2], and some simple models have appeared that explain the observed power law behaviour in financial indices [3,4]. For manufacturers of consumer goods and retailers it could be quite advantageous to apply similar models to sales to the consumer [5]. However, apart from a few exceptions [6,7,8,9] there has been little effort in modelling consumer behaviour via agent-based models.

A credible model should reproduce actual market data. To characterise the consumer goods market, the fluctuations in weekly sales of three related products (ketchup, mayonnaise and curry) were analysed recently [10]. Fluctuations in the baseline sales characterise the dynamics of the market. Some hitherto unnoticed effects were found that are difficult to explain from simple econometric models. In this market the noise level of baseline sales appears to be much larger than can be expected from independent sales events. Further, the sales differences between successive weeks follow a (truncated) Lévy distribution and the auto-correlation of noise decays over a period of some 10 weeks. Finally, a cross-correlation was found between brands, which decays at the same time scale. These effects can be used to test a model.

Several effects may give rise to a power law distribution in sales. One possibility is that the market is in a self-organised critical state, and that "avalanches" occur regularly caused by a Darwinian evolution mechanism [11,12,13]. However, in models of stock markets like the Minority Game, self-organised criticality arises from the competition between agents for scarce assets [1]. This may be realistic for investors, but not for consumers who gain little from being in the minority. In general, the occurrence of a power law in sales returns follows directly from having the right power law distribution of "investor" sizes. If a time series is generated by adding random "superspins" at each time step, the resulting signal is truncated Lévy distributed, provided that the size of superspins is chosen from the correct power law distribution [14,15]. In such a model each superspin represents a cluster of investors whose actions are correlated. How these clusters are formed is irrelevant for the resulting noise spectrum.

Since it was observed that the noise in sales to consumers follow a (truncated) Lévy distribution, it can be concluded that consumer actions are correlated in clusters that follow a power law distribution. The question thus is: what is the cause of this correlation? Why do many consumers take the same decision at the same time? In general the decision of a consumer to purchase a particular product depends on habit, social effects, perceived product quality, price and advertising exposure [16,17,18]. Two factors here may induce correlation between consumers: their being linked in a social network and the stream of advertisements leading to common experiences for large groups of consumers.

To study these two factors, a consumer market is simulated by an agent-based model where all consumers are linked in a social network. The paper is organised as follows. In section 2 the choice model that is studied is described. In section 3 we study the influence of network topology and of the strength of the social interaction. Next, the effect of collective experiences is studied in section 4, and results and conclusions are summarised in section 5.

## 2   Modelling Choice

We simulate the market with a q-state spin model, where $N_b$ = q–1 states are brand choices, and the last choice is to buy nothing. The most commonly used model to represent consumer choice behaviour is the logit model. In this model, choice is represented by a probability P to buy a certain brand b, which depends on the perceived utility $u_b$ of the brand and on its price $p_b$. The probability that consumer *i* will choose brand *b* in the logit model is given by [16, 19, 20, 21]:

$$P(i,b) = \exp[\beta \pi_{ib}] / \sum_{b'=1}^{q} \exp[\beta \pi_{ib'}] \sim \exp[\beta(u_{ib} - p_b)] \qquad 1$$

Utility $u_{ib}$ in general is a measure of what the consumer *i* is prepared to pay for a product *b*, i.e. its perceived value. Parameter β in Eq 1 can be interpreted as miserliness and takes a role analogous to inverse temperature. When β is small, the consumer does not care about price, but rather prefers to seek variety. When β is large the consumer tries to get the best product to the lowest price, and doesn't care about sampling different brands.

In the present investigation we wish to study both the importance of social interactions and of advertisements. To study the former, the perceived utility is split in a quality $q_{ib}$ that is perceived as intrinsic to the brand, and a part that is related to the market share $\psi_{ib}$ of that product over the peer group of $i$. The excess of product utility over its price, as perceived by consumer $i$ is thus written as [17]

$$\pi_{ib} = u_{ib} - p_b = (1-c)(q_{ib} - p_b) + cN_i\psi_{ib} \qquad 2$$

where $N_i$ is number of peers of consumer $i$. The social interaction term represents the number of times that brand $b$ is observed to be chosen. It describes that the market share $\psi_{ib}$ of brand $b$ over the peer group of consumer $i$ leads to an increase of the perceived product value. Parameter c switches between consumers who are independent (c = 0) and those who are connected, communicating and co-operative with their network (c = 1).

According to Becker and Murphy [22] the willingness to pay for a product is increased by advertisements. This implies that when a (positive) news event arrives at a consumer, it raises his/her perception of utility of that brand. To simulate such news events, every consumer $i$ retains a perceived quality $q_{ib}$ of brand $b$, which is an accumulation of his/her news history. At every time step a set $X$ of the population is randomly selected whose perception will be changed. The number of its elements $x = |X|$ is a random variable drawn from a power law distribution given below. Every time step one brand $b'$ is selected at random for which a news item will be broadcasted. The consumers who are selected to be influenced change their perceived value of the selected brand, i.e.

$$q_{ib}(t) = \begin{cases} q_{ib}(t-1) & \text{if } i \notin X \\ q_{ib}(t-1) + \sigma(t)\delta_{b,b'} & \text{if } i \in X \end{cases} \qquad 3$$

For symmetry, the update variable for the perceptions $\sigma(t) = \pm\sigma$ has a randomly selected sign, but it is equal for all consumers in that advertisement round. The absolute value $\sigma$ is a fixed parameter and will be taken equal for all consumers. The interpretation of this parameter is consumer susceptibility to advertisements. It is the added value of an advertisement. After the consumers are 'seeded' by good or bad news, they enter a buying round in which they decide to buy any of the brands or to buy nothing. The perceived quality $q_{ib}$ of brand $b$ is reset to zero when it is bought by consumer $i$.

The source of collective external stimuli captured by Eq (3) could be anything ranging from exposure to advertisements and promotional activities, to the daily news. To get a rough indication of the importance of endogenous influences, buying behaviour under price promotions was analysed for the time series of sales of ketchup, mayonnaise and curry sauce in the Netherlands. The data was extracted from an ACNielsen database [23]. The data available for this work covers a 120-week period. ACNielsen not only gives the actual sales per brand, but also gives separate results for promotion excess sales. Here we analyse these excess sales due to promotions, as this may be an indication of how often people are influenced by external stimuli. We hypothesise that the influence under advertisements follows the same distribution as that of promotions.

Sales excess under promotion was ranked to size $x$. If the probability density to have sales $x$ is given by $\rho(x)$, the ranking number of a promotion is $n(x) = \int_x^\infty \rho(x)dx$, hence the derivative of the ranking number to sales gives the probability density $\rho(x)$. For all products studied we thus find that the frequency of influential events roughly follows

$$\rho(x) \sim x^{-1.06} \qquad \qquad 4$$

Note that this exponent 1.06 is much smaller than the critical percolation exponent $\gamma = 2.5$ that is used in the Cont-Bouchaud model to describe the distribution of investor sizes [14]. Montroll and Shlesinger [24] pointed out that if a random process $P = P_1P_2P_3..P_n$ is the product of many independent random processes, the variable P will be distributed log-normal. This may explain the apparent hyperbolic power law. For some 80% of all promotions the cumulative distribution $n(x)$ is proportional to the logarithm of their size $x$, implying a hyperbolic law for the probability density as in Eq (4). Hence, a power law distribution according to Eq (4) was hypothesised for the impact of *advertisements* and the consequences of such power law are studied.

The simulation contains $N_b$ brands and $N_c$ consumers. At each time step every consumer $i$ makes only *one* choice $b$ out of the set {none, brand 1, …, brand $N_b$} leaving out the previous choice. The previous choice can be either a brand, or not to buy anything. Whether this switch is accepted or the previous choice is retained, depends on whether the perceived value, $\pi_{ib}$ goes up or down relative to the current perceived value of the latest choice. If $\pi_{ib}$ increases, the new choice is accepted immediately, but if $\pi_{ib}$ decreases the new choice is accepted with probability

$$P(accept) = \exp[\beta(\pi_{ib}^{new} - \pi_{ib}^{old})] \qquad \qquad 5$$

which is the standard Metropolis Monte carlo algorithm. To set the fraction of non-buyers in the system, each consumer has a personal quality standard. Every time a product is consumed, the standard is increased by a random number taken from (0, $\delta$), where $\delta$ = 0.5($N_c$–$M$)/$N_c$ and every time the consumer refrains from consuming his standard is decreased by an amount (–$\varepsilon$, 0), where $\varepsilon$ = 0.5$M$/$N_c$. This mechanism self-adjusts to a desired market volume $M$, even though the total market volume is not fixed. The quality

RD Groot, Social networks and advertisement effects in consumer choicereference performs the function of defining the utility (or chemical potential) of not buying, i.e. it is substituted in Eq 5 for $\pi_{ib}^{new}$ if the new choice is "none".

Since the utility of products in this model is partly imputed via their use in the peer group of a consumer, all consumers are placed in a network. As it is not apparent which network topology is pertinent for consumer interactions, some choices have been studied. These are the small-world network [25,26], the Barabási-Albert network [27] and a random network. Results for these networks are described in the next section.

## 3   Sales without advertisements

In these simulations 3000 consumers are grouped in a circle, where each agent is connected to four nearest neighbours. A random fraction of 3% of the connections are broken, and rewired to a randomly chosen other agent. Hence, the network can be characterised as a 1+ε dimensional small-world network. We arbitrarily put the inverse temperature at β = 2, and vary the value of the consumer co-operativity c. Prices are taken as fixed random numbers between $p_b$ = 0 and $p_b$ = 1. The market contains $N_b$ = 10 brands. The actual chosen set of prices is {$p_b$} = {0.119, 0.161, 0.289, 0.497, 0.524, 0.631, 0.747, 0.839, 0.962, 0.975}.

In figure 1a the evolution of sales is shown for c = 0. All consumers act independently from the others. Figure 1b shows a simulation on the same 3000 virtual consumers, and on the same brands, but now the consumer perception of the product value is for 40% determined by the price of the product, and for 60% it is determined by what the others buy (c = 0.6). A few top brands are established that do well, whereas all other brands are strongly reduced in market share. Moreover, by introducing social utility, the fluctuations in the top brand market share are largely increased. Such fluctuations are usually seen in real market share data [10]. Finally in Fig 1c an example is given where c = 0.75, in which the top-brand gains nearly complete market dominance. At roughly c = 0.7 the system has a transition point that *resembles* a phase transition, where it shows a maximum in volatility, see Figure 2.

At this maximum, the amplitude of sales fluctuations is some 10% of sales for the top brand. This is realistic compared to sales in the ketchup market. However, when we study the noise distribution of sales, we find a marked difference between the real sales data to consumers and the present simulations at the "critical" point, see Figure 3. Whereas empirical sales data shows a power law distribution of fluctuations, closely resembling a truncated Lévy distribution, the simulation on the present small-world network has Gaussian fluctuations.

Increasing the number of consumer by a factor of 10 has no influence on the shape of the noise spectrum, it remains Gaussian. This means that the model may have a realistic noise *amplitude*, but the large peaks in sales that characterise the consumer data are missing. The implication is that the underlying model is not a realistic model for the way consumers operate, or for the way they are placed in a network.

The situation on a 2+ε dimensional small-world network, which is made by rewiring 3% of the connections of a 2D network, are qualitatively similar. The volatility at the critical point has a sharper maximum as function of c, but the fluctuations at the critical point still follow a *Gaussian* distribution. Also when the buying frequency is lowered a factor of ten, Gaussian noise is found. Simulations have also been done on a random network with a stretched exponential connectivity distribution. The network was built by rewiring a Barabási-Albert network, using a Verlet neighbour list [28] (see Appendix). The same set of brands was used as before. This network shows mean-field behaviour, with a strong first order phase transition and hysteresis behaviour, see Figure 4. The drawn curves are mean-field results, obtained from solving

$$\psi_b = \frac{\exp[\beta(4c\psi_b - (1-c)p_b)]}{\sum_{b'} \exp[\beta(4c\psi_{b'} - (1-c)p_{b'})]} \qquad 6$$

where the number of neighbours per agent was set on its mean value $N_i = 4$.

In the thermodynamic limit the Ising model on the small-world network [29] shows mean-field behaviour with classical exponents at the critical point for every rewiring fraction. Since the underlying reason this is the topology of the small-world network [30], the same must hold for the q-state spin model. This *excludes* the present q-state choice model on any small-world network as describing consumer choice behaviour, which indicates that some crucial factor must be missing in this model. Moreover, this model shows critical slowing down at the critical point, which is not seen in supermarket sales.

Sales fluctuations of consumer products might be (truncated) Lévy distributed because the network of consumer relations is strongly inhomogeneous. Therefore, we also studied sales on a directed Barabási-Albert network [27]. Here the probability for any node in the network of having k connections to others decays as

$$P(k) \propto k^{-\gamma} \qquad 7$$

When the number of connections between consumers follows a power law, this implies that also the influence of any person on all others in the network follows a power law. This may have profound influence on buying behaviour, as the ones who have many connections act as *key opinion formers*. Indeed, buying behaviour on this network is markedly different from that on a small-world network. Simulations were run with the same product prices as before, and $\beta = 2$. Again a network of 3000 agents was generated with connectivity m = 4 (see Appendix). Agents only take advice from their $N_i$ = m = 4 outgoing links.

In Figure 5 consumer co-operativity is increased from the left to the right, it varies as c = 0.1, 0.5 and 0.7. For low co-operativity the market shows few fluctuations, like on the small world network. When c = 0.5, the buying behaviour is very chaotic. Short fashions can be distinguished, but the market is never stable. If consumers are highly co-operative, the market is characterised by longer fashions, where nearly all consumers are converted to the same brand. The main effect of increasing c above c = 0.5 is to increase the time for which a certain fashion lasts. All markets for c > 0.5 demonstrate this behaviour, which resembles a glassy state. The system does not show phase separation behaviour like on the small-world network, instead the system swaps between different product fashions, where typically 98% of the consumers buy the same product. After a certain period of time, all consumers switch to another product. Examples of such transitions are shown in Figure 5c. This phenomenon is entirely driven by herding behaviour, price and quality are irrelevant as transitions occur from cheap to expensive brands and vice versa.

The distributions of sales difference between successive measurements are shown in Figure 6 for systems of c = 0.3, 0.5 and 0.7. The dots in this figure are obtained from real weekly sales differences of ketchup, mayonnaise and curry sauce to Dutch consumers, averaged over a period of 118 weeks [10]. The simulated distributions again show a marked difference for c < 0.5 and c > 0.5. For co-operativity below c = 0.5 the noise distribution roughly follows a stretched exponential decay, $P(x) \propto \exp(-x^\alpha)$ where $0.7 < \alpha < 2.0$. At c = 0.5 it follows a (truncated) Lévy distribution, crossing over to a steeper power-law decay, $P(x) \sim x^{-3.7}$, probably due to finite size effects. For c > 0.5 the distribution shows a fast initial decay, then a plateau, and finally a power law decay. The point c = 0.5 appears to be close to a transition point separating a glassy system from a non-glassy system. This transition point is markedly non-classical, given the broad noise distribution. Without any fitting parameters, the shape of the distribution at the transition point matches the noise distribution seen in consumer sales data. However, the noise amplitude is much higher than what is seen in a typical consumer market, some 80% of the market share for the top three brands, whereas this is 8 to 12 % in the supermarket [10]. The noise amplitude in the simulation is hardly dependent on the number of consumers in the simulation, or on the time span over which sales are averaged for each measurement. There is apparently no parameter set in this model that matches all empirical data.

## 4  Sales with advertisements

In this section the present market model will be compared to actual market data, to see if it is possible to reproduce the noise spectrum, noise amplitude and correlation functions as appear in a real market. To this end we simulate the Dutch ketchup market, where four brands are active with a total sales of about 71000 bottles a week. Simulations were done in the parameter range between c = 0 and c = 0.5. The market size M was put at an average of M = 11800 bottles per day, and sales over six days were added to obtain weekly sales. A good or bad news item about one brand is disseminated daily. For each value of the consumer co-operativity, price variables were fixed to reproduce the market shares of four brands (see Table I). For all simulations miserliness was taken as $\beta$ = 2, and the marketing susceptibility $\sigma$ was chosen to match the sales auto-correlation function. Simulations were run on a (1+$\epsilon$) dimensional small-world network with 3% rewiring and on average $N_i$ = 4 neighbours per agent.

Generally, power law noise only occurs when the market is sufficiently large, otherwise the rare extreme events are not sampled adequately [31,32]. Therefore we first check on finite size effects. In these simulations the social co-operativity was fixed at c = 0.3, the noise amplitude was taken as $\sigma$ = 0.173, and $N_c$ was varied as $N_c$ = 465, 4650, 15500 and 46500 agents and for each market M = 0.254$N_c$ bottles per day. At this point the simulated market size equals that of the real market, though the buying frequency is unrealistic. The distributions of sales differences between consecutive weeks for these runs are shown in Figure 7. For comparison, the dashed curve is a Lévy distribution of exponent $\alpha$ = 1.4. The actual noise distribution obtained from ketchup, mayonnaise and curry sauce markets is shown by the full dots with error bars. An excellent match with the empirical data is obtained. At the prices {0.0, 0.219, 0.319, 0.372} we find the market shares and fluctuation amplitudes, in simulation and in the real market as given in Table I. The shape of the noise distribution follows naturally from the power law assumed in the promotional effectiveness. When the power –1 in Eq 7 is replaced by the percolation exponent –2.5, as in the Cont-Bouchaud model [14], the noise distribution is too narrow to fit the market.

To further compare the simulations to market data we study the correlation functions:

RD Groot, Social networks and advertisement effects in consumer choice$$C_{ij}(\tau) = \frac{\langle (s_i(t) - \bar{s}_i)(s_j(t+\tau) - \bar{s}_j) \rangle}{\sqrt{\bar{s}_i \bar{s}_j}} \qquad 8$$

where $\bar{s}_i$ is a time average of sales $s_i(t)$ over the interval $0 < t < T-\tau$, and $\bar{s}_j$ is a time average of sales $s_j(t)$ over the interval $\tau < t < T$. The correlation function depends on whether or not the empirical data is detrended. By analysing simulated time series over 118-week periods with and without detrending the data, the correct correlation functions can be estimated reasonably well by taking a linear combination of detrended and non-detrended correlation functions, such that the correlation function vanishes after 30 weeks. This procedure was subsequently used to analyse the empirical market data. The results are shown in Figure 8, and are compared with the simulated correlation functions. At the parameter value where the market dynamics for ketchup sales is reproduced, social effects determine product utility for roughly 50%.

Social effect particularly influence the cross-correlation between the three lower brands. If we take c = 0, the cross-correlation functions in the simulation are negative, whereas this is clearly positive in the market. To arrive at the same fluctuation amplitude without social interaction, the noise level was increased to $\sigma = 0.22$ and the number of consumers was taken as $N_c$ = 48000. For these parameters the auto-correlation decays with a power law of time, as $C(t) \sim t^{-0.78}$, see Figure 9. This result is based on a run of $5 \times 10^5$ time steps after an equilibration over 6000 steps. Even though this run is quite long, there is still a slight difference between the correlation functions obtained from detrending the data and not detrending. In the latter case a slightly less negative power $C(t) \sim t^{-0.68}$ is obtained. In the present simulations, big (exogenous) shocks must dominate the correlation function, therefore we should expect [33] $C(t) \sim t^{-(1-\theta)}$ from which we find $\theta = 0.27 \pm 0.05$. This result is identical to the result obtained by Sornette *et al* for exogenous shocks in book sales [34]. Further increasing the number of consumers and noise amplitude to $N_c$ = 52840 and $\sigma$ = 0.28 leads to a power law correlation $C(t) \sim t^{-0.4}$, see Figure 9. For these parameters the noise amplitude is about twice as high as in the real ketchup market. Remarkably, this power law is identical to the result obtained by Sornette *et al* for *endogenous* shocks [34], but this correspondence may be fortuitous.

In the present model, long time correlation develops because a fraction of consumers develops a particularly negative view of some brands. This can be studied by making a histogram of the number of times a particular value for the perceived product quality $q_{ib}$ occurs. This is averaged over all brands for each consumer and shown in Figure 10. When the agents ignore the behaviour of others, this is their actual perception of a brand. For a brand that is perceived as very negative by some agents, buying probability decreases exponentially as P($i,b$) ∝ exp($\beta q_{ib}$), hence those agents avoid buying these brands. Consequently, their negative views are never corrected. Only when by chance their perceived quality diffuses back to less negative values and they happen to buy the product, their perception is reset. These persistent negative views apparently lead to long time correlations. When the agents do value the behaviour of others (c = 0.3), brand quality is also determined by what the neighbours do. As a result, very negative prejudices have a much smaller tendency to develop. In the latter case no power law correlation develops (as in Figure 9), but the correlation decays exponentially, see Figure 8.

## 5  Summary

In this paper the importance of social interaction between consumers and of collective marketing effects have been studied by simulation of the choice process. Agents were placed in a small world network, a random network and in a scale-free network. They interact with their peers by exchanging information on the latest brand choice. To study the effect of advertisements and other common experiences amongst the consumers, a model is introduced where advertisements lead to collective imprints that ripen at the decision point. The simulations are compared to sales data of ketchup on the Dutch market.

On the small-world network a transition is found between lock-in and an open market. In a lock-in situation one market leader dominates the market and all other brands remain small. This occurs when consumers tend to copy the behaviour of their neighbours and want the best value for their money. This is the market analogue of condensation in liquids. An open market arises when consumers ignore what their friends buy. At the transition between these two extremes, fluctuations in the market become large and buying behaviour becomes strongly correlated. However, the sales fluctuations seen in the real market *cannot* have their origin in this criticality for two reasons. Firstly, the critical point on a small-world network as studied here has Gaussian fluctuations. Secondly, at the critical point the simulations show critical slowing down, which is not seen in consumer markets.

On the Barabási-Albert network, people tend to follow the behaviour of a few key opinion formers. When this influence is strong the market drops into a glassy state where the market leader attains nearly complete market dominance. In this state fashions alternate. When consumers put little weight on the opinion of their peers, products are chosen according to the logit distribution of their price. In between, the system shows critical behaviour with fluctuations well characterised by a truncated Lévy distribution. However, the amplitude of the fluctuations is much higher than what is seen in the market, and moreover the noise distribution spectrum could *only* be reproduced when the parameters of the model are fine-tuned to a particular value. Hence the broad fluctuation spectrum seen in supermarket sales is neither caused by near neighbour interactions between consumers, nor by the network topology in which they are connected. For this reason, exogenous causes for the observed sales noise spectrum are considered. A likely cause for correlation between consumers is collective imprints by advertisements.

The distribution of sales under promotion is taken as a model for collective experiences that change behaviour. The probability of such events is roughly inversely proportional to the size of the event. This apparent hyperbolic law might be explained by the fact that each advertisement event only influences a particular consumer when a large chain of independent switches is satisfied. When this power-law of collective experiences is put into the model, the correct distribution of weekly sales returns is reproduced directly. The correct power law is obtained for the weekly sales returns. At the parameter value where the market dynamics for ketchup sales is reproduced, social effects in this model determine product utility for roughly 50%. This correspondence does not prove that the model is correct, but so far it is consistent with the empirical data.

Even though this is a very simple model of the mind, it leads to a very rich phenomenology. A remarkable result is obtained when the distribution of perceptions is studied. Positive prejudices towards a brand are very short lived, as the agents immediately buy these products and therewith readjust their perception. Negative prejudices however can be very long lived, as the agents avoid those products. Persistent negative views particularly build up when agents operate in isolation, but when they take into account the behaviour of their peers to value a product, the development of extremely negative views is prevented. The reason for this is that copying peer behaviour leads to a frequent exposure to the product, which corrects negative prejudices. Thus the model predicts that social interaction can prevent the development of extreme views, that aversion is more often developed when the buying frequency is lower, and that aversion is developed towards the unknown. When consumers do not value the buying behaviour of their peers, sales auto-correlation and cross-correlation develop power laws of time, that depend on the parameter values. Powers observed in the simulation are identical to the power laws observed in the decay of endogenous and exogenous shocks in book sales [34]. The appearance of power laws suggests a self-organised critical state, like in the avalanche model [11].

## Appendix. A fast algorithm to generate a Barabási-Albert network

The Barabási-Albert network [27] is formed by adding new nodes one by one to an existing network. Each new node is connected to $m$ existing nodes. However, the probability to connect to a particular node depends on the number of connections that the latter already has. If node $i$ is connected to the network by $k_i$ links then the probability that $j$ connects to $i$ is given by

$$P(i) = \frac{k_i}{\sum_{l<j} k_l} \qquad \text{A1}$$

In social terms, people will listen preferentially to influential people. A fast algorithm to generate such a network is the following. Each person $i$ in an existing network has $m$ outgoing connections and $k_i - m$ incoming connections (persons listening to $i$). To add a new person $n$ to an existing network, take any randomly chosen person $i$ and consider randomly one of his/her $m$ outgoing links connecting $i$ to $j$ ($i$ takes advice of $j$). With probability 0.5 connect $n$ to $i$, otherwise connect $n$ to $j$.

This procedure leads to the correct distribution given by Eq A1, because all links are sampled with equal probability. The probability to arrive at a particular person is therefore proportional to the number of departing links $m$ plus the number of arriving links $k_i - m$. Since both ends of the link are taken with equal probability the total probability to choose a particular person is proportional to $k_i$. This procedure solves the obvious problem: how do people know how often anybody is connected to the network? Without citation index they don't know. An obvious generalisation of this procedure is to follow the advice of the first arbitrary person with probability f (i.e. connect to $j$) and connect to $i$ himself with probability (1–f). For f = 0.5 this leads to the original Barabási-Albert network with exponent $\gamma$ = 3.0 when the number of nodes is large ($10^5$-$10^6$). For f = 0.7 a slightly lower exponent $\gamma$ = 2.4 is found, which is close to the actor network.

When rewiring is introduced in a fully-grown network, the topology changes significantly. To model this, all agents die in a random sequence. When an agent $i$ dies, its connections are broken, and all $k_i - m$ bereaved agents find a new connection according to the procedure described above. Finally the deceased agent is reborn and chooses new connections to the network. When this procedure is repeated for some generations a stable distribution of the number of nodes per link emerges. For this network the distribution of connections is no longer algebraically decaying, but it appears to decay as a stretched exponential,

$$N(k) \approx 4.5\exp(-1.5k^{0.54}) \tag{A2}$$

More in general, rewiring by birth-and-death increases the exponent $\gamma$ in Eq 9, or for small f it destroys the power-law behaviour altogether.

The network can also be rewired while it is being grown. In that case, at every step a choice is made between adding a new node with probability (1–r) or rewiring an existing node with probability r. To this end, a node $i$ is selected randomly from the nodes that have been defined already, and this node is given new outgoing connections while the incoming connections remain intact. Generating a network with rewire probability r = 0.5 appears to lead to the same network as obtained after many generations of death and birth, in which both outgoing and incoming connections are broken.

To keep tack of all links we use a Verlet neighbour list, as commonly used in the simulation of liquids [28]. This is however not necessary to generate the initial network, it is only necessary if the grown network is to be rewired afterwards using the birth-and-death algorithm. To grow the initial network the outgoing links are stored in an array out[$N_c$][m], and for every agent $i$ we remember how many other agents N[$i$] point to $i$. This is enough information to form the network, since we only follow outgoing links to form the network.

After the initial network is generated, the neighbour list is generated. This is an array neighbour[m$N_c$] of size m$N_c$ that contains the label numbers of the incoming neighbours that point to any node. The neighbours of $i$ are stored in the elements neighbour[pointer[$i$]] up to neighbour[pointer[$i$]+N[$i$]–1], where pointer[$i$] = $\Sigma_{j<i}$ N[$j$] points to the begin point where the information for node $i$ is stored. The array is filled by first setting up these pointers, then resetting the connection array N[$i$], and finally running through the outgoing connections. If $j$ connects to $i$ = out[$j$][$k$], we add 1 to N[$i$], and store $j$ in neighbour[pointer[$i$]+N[$i$]-1]. When all outgoing links are checked this way, the connection array N[$i$] is restored again.

## Table and Figures

**Table I**. Summary of actual and simulated market shares and fluctuation amplitudes.

**Figure 1** Market shares for ten brands of randomly chosen price at c = 0 (left, all consumers act independently), c = 0.6 (middle), and c = 0.75 (right, product value is for 75% determined by market share under peers).

**Figure 2** Market share of ten brands as function of the consumer co-operativity c, their propensity to follow their peers. Low c means consumers are independent, high c means consumers are connected and co-operative. The dashed curve is the relative fluctuation amplitude of the market leader.

**Figure 3** Noise distribution of sales returns for consumer data (o) and simulation on small-world network ($\Delta$) at its critical point.

**Figure 4** Market share of ten brands as function of the consumer co-operativity c on a random network. Brand prices are the same as in Figure 2. Curves are based on mean-field theory, Eq 6. Note the hysteresis behaviour, indicating a strong first order phase transition.

**Figure 5** Typical patterns of buying behaviour on Barabási-Albert network for consumer co-operativity c = 0.1, 0.5 and 0.7, at $\beta$ = 2.

**Figure 6** Sales difference distribution on a Barabási-Albert network for c = 0.3, 0.5 (full curve) and 0.7. The dots are results from consumer sales data.

**Figure 7** Noise distribution in simulated sales returns for 465 (squares), 4650, 15500 and 46500 (circles) agents, showing a gradual crossover to a truncated power law distribution as the numbers of agents increases. Other simulation parameters are: c = 0.3 and $\sigma$ = 0.173. Actual market data are indicated by full dots with error bars.

**Figure 8** Sales correlation function of consumer data (dots) and simulation results (lines). Symbols denote: circular dots, auto-correlation; triangles, top-low cross-correlation; squares: low-low cross-correlation. Simulation parameters are: c = 0.3, $\sigma$ = 0.173, and $N_c$ = 46500.

**Figure 9** Sales auto-correlation function without social interaction between consumers (c = 0), for simulation parameters $N_c$ = 48000, $\sigma$ = 0.22 (slope –0.75) and for $N_c$ = 52840, $\sigma$ = 0.28 (slope –0.4).

**Figure 10** Distribution of consumer held prejudices $q_{ib}$, averaged over all brands, for two values of social co-operativity. Parameters are c = 0, $N_c$ = 48000, $\sigma$ = 0.22; and c = 0.3, $N_c$ = 46500, $\sigma$ = 0.173.

| Brand | Market shares | | Relative fluctuation amplitude | |
|---|---|---|---|---|
| | real market | simulation | real market | simulation |
| Heinz | 0.527 | 0.534 | 0.067 | 0.077 |
| Calve | 0.196 | 0.195 | 0.110 | 0.119 |
| Gouda's Glorie | 0.148 | 0.145 | 0.081 | 0.122 |
| Remia | 0.129 | 0.126 | 0.124 | 0.125 |

**Table I**.

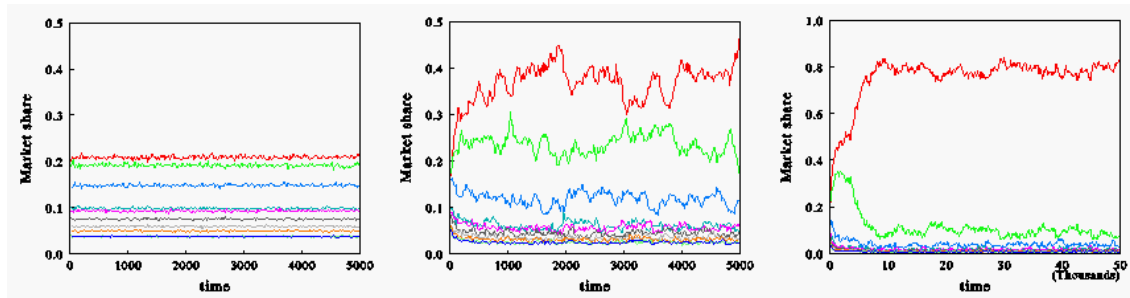

**Figure 1**

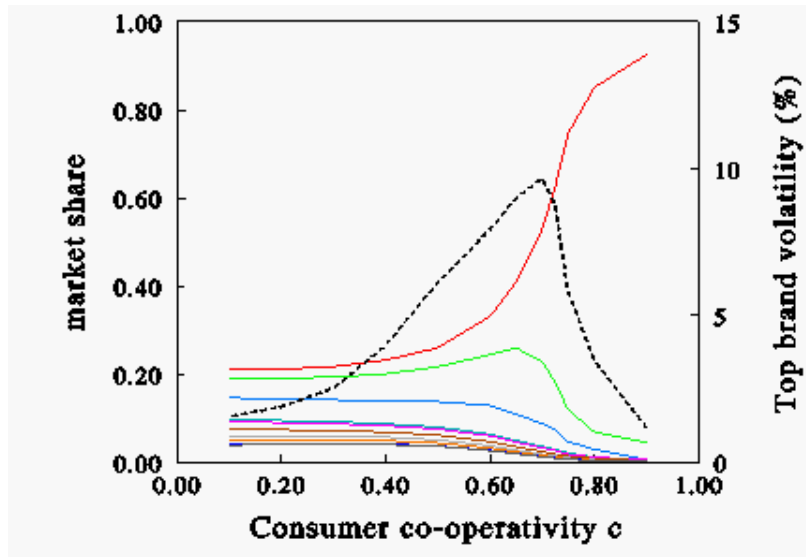

**Figure 2**

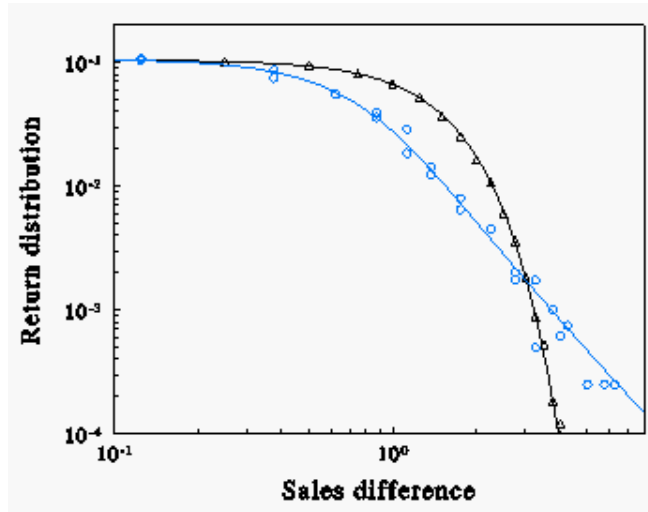

**Figure 3**

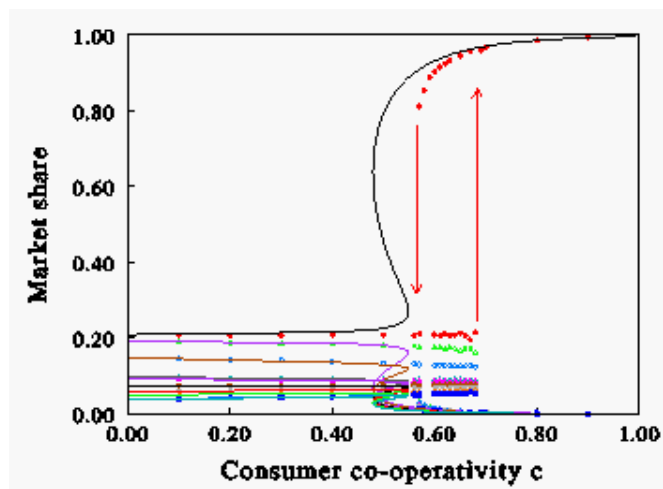

**Figure 4**

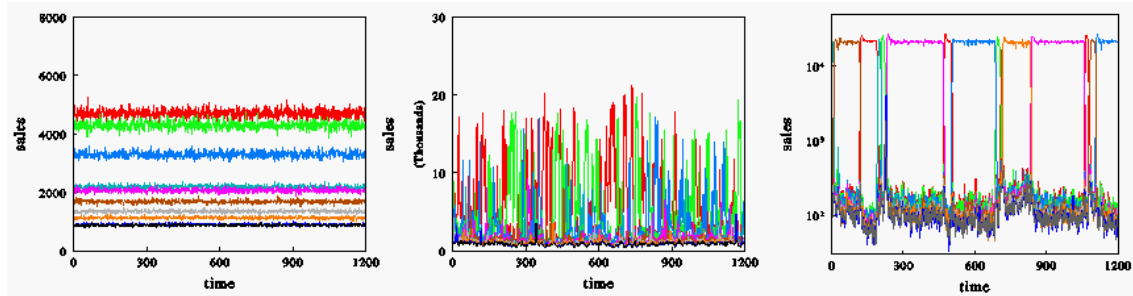

**Figure 5**

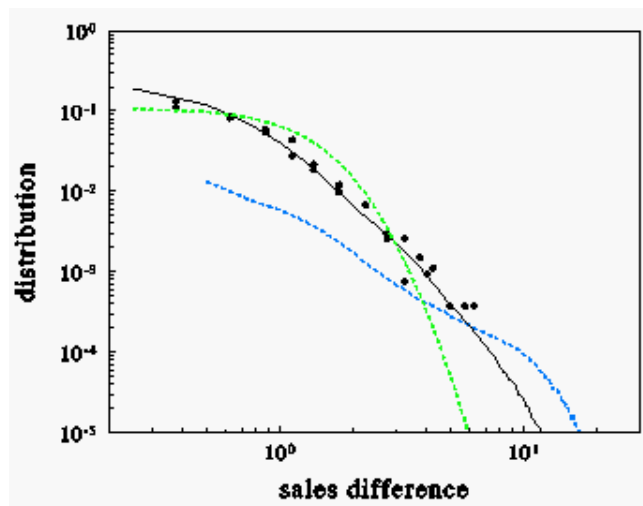

**Figure 6**

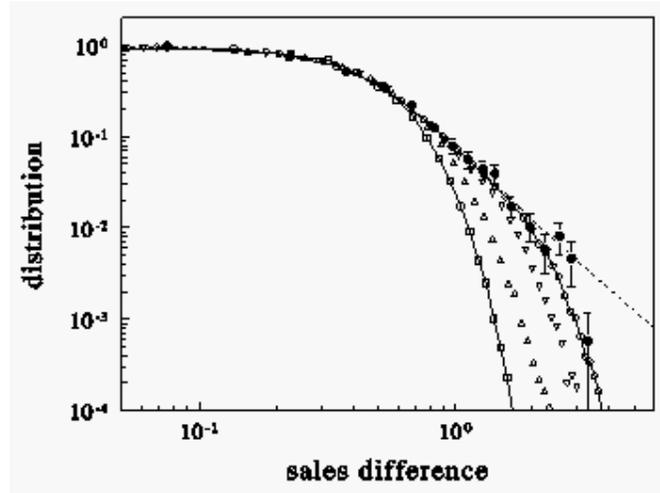

**Figure 7**

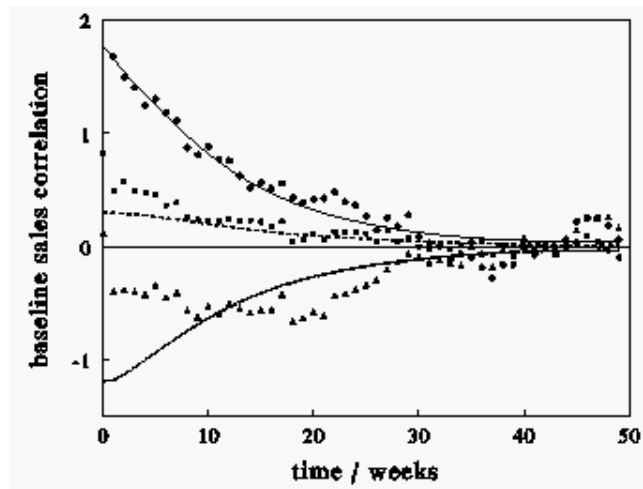

**Figure 8**

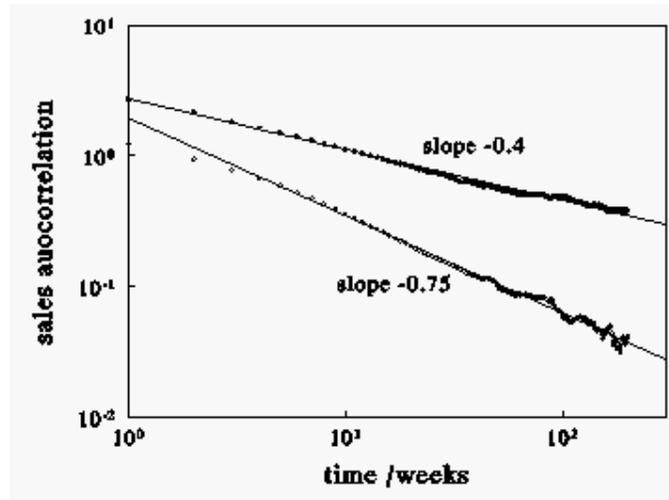

**Figure 9**

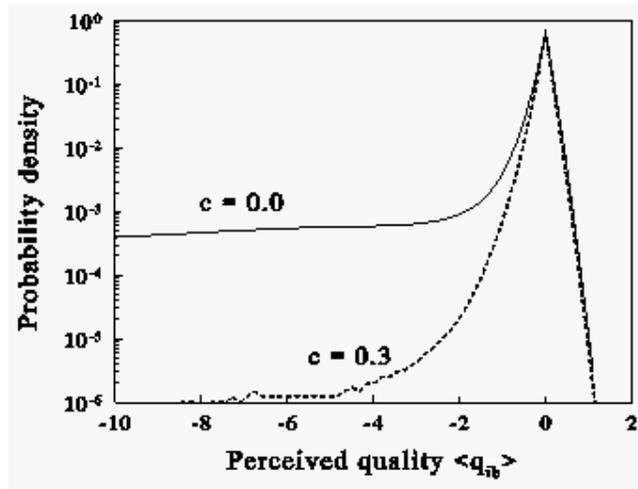

**Figure 10**